\documentstyle[preprint,aps,epsfig]{revtex}
\tightenlines

\begin{document}

\preprint{ }
 
\title{$J/\psi$~ Suppression and Enhancement 
\\ in Au+Au Collisions at the BNL RHIC}

\author{ 
M.I. Gorenstein$^{a,b}$,
A.P. Kostyuk$^{a,b}$,
H. St\"ocker$^{a}$
 and  
W. Greiner$^{a}$
}

\address{
$^a$ Institut f\"ur Theoretische Physik, Goethe Universit\"at,  
Frankfurt am Main, Germany}

\address{$^b$ Bogolyubov Institute for Theoretical Physics,
Kyiv, Ukraine}

\maketitle

\begin{abstract}
We consider the production of the $J/\psi$ mesons in heavy ion collisions
at RHIC energies in the statistical coalescence model
with an exact (canonical ensemble) charm conservation. The 
$c\overline{c}$ quark pairs are assumed to be created in the 
primary hard parton collisions, but the formation of the
open and hidden charm particles takes place at the 
hadronization stage and follows the prescription of statistical mechanics.
The dependence of the $J/\psi$ production on both the number of
nucleon participants and the collision energy is studied.
The model predicts the $J/\psi$ {\it suppression} for low energies,
whereas at the highest RHIC energy the model reveals the $J/\psi$
{\it enhancement}.
\end{abstract}


The experimental program for studies of the charmonium production 
in nucleus--nucleus (A+A) collisions at CERN SPS over
the last 15 year was mainly motivated
by a suggestion of Matsui and Satz \cite{MS} to use the 
$J/\psi$ as a probe of the state of matter in the early stage
of the collision. The original picture \cite{MS} 
(see also Ref.\cite{Satz} for a modern review)
assumes that charmonia are produced in primary collisions
of the nucleons from the colliding nuclei. The number 
of created charmonium states 
is then reduced because of inelastic interactions 
with the nucleons of the colliding nuclei.
An additional suppression may occur due to $J/\psi$ interaction with 
the secondary hadrons (`co-movers') \cite{comover}. 
The probability to  destroy the charmonium state increases obviously
with  the number of nucleon participants
$N_p$. Similar behavior is expected when the collision energy
$\sqrt{s}$ increases because the number of produced
hadrons (`co-movers') becomes larger. 
This is known as the {\it normal $J/\psi$ suppression}.
 Furthermore, the $J/\psi$ 
(and other charmonia) are assumed to be formed mainly from the
$c\overline{c}$ pairs with invariant mass below the 
$D$-meson threshold \cite{Gavai}.
The fraction of these subthreshold pairs in the total 
number $N_{c\overline{c}}^{dir}$ of
$c\overline{c}$ pairs (which is roughly
proportional to the number of produced open charm hadrons
($D$,  $D^*$, $\Lambda_c$ etc.) also decreases with
$\sqrt{s}$. Therefore, the ratio
\begin{equation}\label{ratio1}
R(N_p,\sqrt{s})~\equiv~\frac{\langle J/\psi \rangle}
{N_{c\overline{c}}^{dir}}
\end{equation}
is expected to decrease with increasing  $N_p$ and/or $\sqrt{s}$. 
At large values
of $\sqrt{s}$ and $N_p$ the formation of the quark-gluon plasma (QGP)
is expected which is supposed to be signaled by the {\it anomalous
suppression} \cite{Satz} of the $J/\psi$, i.e. a sudden and strong
decrease of the ratio (\ref{ratio1}) is considered as a signal of the QGP
formation. Hence, a decrease of the ratio (\ref{ratio1})
is an unambiguous consequence of the standard picture \cite{Satz,comover}.

A very different approach of the statistical $J/\psi$ production, 
proposed in  Ref.~\cite{GG}, assumes that
$J/\psi$ mesons are created at the hadronization stage similar 
to other (lighter) hadrons. 

A picture of the $J/\psi$ creation via
$c$ and $\overline{c}$  coalescence 
(recombination) was subsequently developed within  different
model formulations \cite{Ka:00,Br1,Go:00,Le:00,Ra:00,Mc:00}. 
Similar to the suggestion of Ref.~\cite{GG}, charmonium states are 
assumed to be created at the hadronization stage of the reaction,
but they are formed due to the coalescence of $c$ and $\overline{c}$,
which were produced by primary hard parton collisions at the initial
stage.

In this paper the  $N_p$ and
$\sqrt{s}$ dependences of the ratio (\ref{ratio1}) will be studied for
Au+Au collisions at RHIC energies.
We use the canonical
ensemble (c.e.) formulation
of the statistical coalescence model (SCM) \cite{Go:00,Mc:00}. 
The number $N_{c\overline{c}}^{dir}$ of the produced $c\overline{c}$
pairs, which is the input
for the SCM calculations, will be estimated within 
the perturbative QCD (pQCD).
The considered pQCD+SCM approach reveals both the $J/\psi$ {\it
suppression} (at $N_{c\overline{c}} < 1$) and the
$J/\psi$ {\it enhancement} (at $N_{c\overline{c}} > 1$)
effects.

\vspace{0.3cm}
In the framework of the ideal hadron gas (HG) model in the grand 
canonical ensemble (g.c.e.) formulation the hadron 
multiplicities are given by
\begin{equation}\label{Nj}
N_j ~=~ \frac{d_j~V}{2\pi^2}~\int_0^{\infty}k^2dk~
\left[~\exp\left(\frac{\sqrt{m_j^2+k^2}~-~\mu_j}{T}\right)~\pm
~1~\right]^{-1}~,
\end{equation}
where $V$ and $T$ correspond to the volume\footnote{
To avoid complications we neglect the excluded volume corrections.
The thermodynamical consistent way to treat the excluded volume 
effects was suggested in Ref.~\cite{Ri} (see also \cite{YG} for 
further details).
If the excluded volume parameter is the same for all hadrons
its effect is reduced only to the rescaling of the volume $V$:
all particle number ratios remain the same as in the ideal hadron gas.
}
and temperature of the HG, $m_j$ and $d_j$ 
denote particle masses and degeneracy
factors. 
Eq.(\ref{Nj}) describes the quantum HG: Bose and Fermi
distributions for mesons and (anti)baryons, respectively.
Quantum effects, however, are found to be noticeable
for pions only, so that Eq.(\ref{Nj}) for all other hadrons
can be simplified to the Boltzmann result:
\begin{equation}\label{Nj1}
N_j ~=~
\frac{d_j}{2\pi^2}~V~
\exp\left(\frac{\mu_j}{T}\right)~
~T~m_j^2~K_2\left(\frac{m_j}{T}\right)~,
\end{equation}
where $K_2$ is the modified Bessel function.

In the case of complete chemical equilibrium the chemical
potential
$\mu_j$ in Eq.(\ref{Nj}) is defined as
\begin{equation}\label{muj}
\mu_j ~=~b_j\mu_B~+
q_j\mu_Q~+~s_j\mu_S~+~c_j\mu_C~,
\end{equation}
where $b_j,q_j,s_j,c_j$ denote the baryonic number, electric charge,
strangeness and charm of hadron $j$. The baryonic chemical potential
regulates the non-zero (positive) baryonic density of the HG
system created in A+A collision. The chemical potentials $\mu_S$ and
$\mu_C$
should be found as the functions of $T$ and $\mu_B$
from the requirements of zero value 
for the total strangeness and charm in the system,
and the chemical potential $\mu_Q$  from the requirement
of the fixed ratio of the electric charge to the baryonic number
(this ratio is defined by the numbers of protons and neutrons in the
colliding nuclei).

The applications of the HG model to fitting the hadron abundances
in particle and nuclear collisions revealed, however, 
a  deviation of strange hadron multiplicities
from the complete chemical equilibrium \cite{Ra:91}. It was 
suggested that strange quarks and antiquarks are distributed 
inside hadrons according to the laws of HG equilibrium, but the total
number of strange quarks and antiquarks inside the hadrons is smaller 
than that in the equilibrium HG  
and remains approximately constant during the lifetime of the 
HG phase.
Therefore, not only the "strange charge"
$N_s - N_{\overline{s}}= 0$ but also the "total strangeness"  
$N_s+N_{\overline{s}}$ is then considered as a conserved quantity.
In the language of thermodynamics, it means an introduction
of an additional chemical potential, $\mu_{|S|}$,
which regulates
this new "conserved" number $N_s+N_{\overline{s}}$. 
Then the additional term, $(n_s^j +n^j_{\overline{s}})\mu_{|S|}$,
should appear in the expression (\ref{muj}) for $\mu_j$ , where
$n_s^j$ and  $n^j_{\overline{s}}$ are the numbers of strange
quarks and antiquarks inside hadron $j$. Introducing a notation,
$\gamma_s \equiv \exp(\mu_{|S|}/T)$ \cite{Ra:91}, one can implement
this additional conservation according to the following simple rule:
the hadron multiplicities $N_j$ (\ref{Nj}) are multiplied 
by a factor $\gamma_s^{(n_s^j +n^j_{\overline{s}})}$, e.g.
factor $\gamma_s$  appears for $K,\overline{K}, \Lambda,
\overline{\Lambda},\Sigma,\overline{\Sigma}$, factor 
$\gamma_s^2$  for $\Xi, \overline{\Xi}$ and factor $\gamma_s^3$ 
for $\Omega,\overline{\Omega}$. For mesons 
with hidden strangeness, like $\eta,\eta',\omega,\phi$, having the
wave function of the form 
\begin{equation}
C_u |u\overline{u}\rangle +
C_d |d\overline{d}\rangle +
C_s |s\overline{s}\rangle
\end{equation}
the factor $\gamma_s^{2|C_s|^2}$  is used.

 From fitting the data on the hadron yield in particle and nuclear
collisions it was found that $\gamma_s \le 1$ for all known cases.
Parameter $\gamma_s$ is called therefore the strangeness
suppression factor.

Recently an analogous procedure was suggested for charm hadrons
\cite{Br1}.
A new parameter $\gamma_c$ has been introduced
to treat simultaneously both the open and hidden
charm particles within statistical mechanics HG formulation. 
The multiplicities $N_j$ (\ref{Nj}) of single open charm and anticharm
hadrons should be multiplied by the factor $\gamma_c$ and charmonium 
states by the factor $\gamma_c^2$. In contrast
to the suppression of strangeness in the HG 
($\gamma_s \le 1$) one observes the enhancement of charm hadron
yields in comparison to their equilibrium HG values.
It means that $\gamma_c \ge 1$ and 
this parameter is called the charm enhancement
factor \cite{Br1}.

To take into account the requirement of zero "charm charge" of 
the HG in the exact form the c.e. formulation was suggested in
Ref.\cite{Go:00}. In the c.e. formulation 
the charmonium multiplicities are still given by Eq.(\ref{Nj1}) as
charmonium states have zero charm charge. The multiplicities 
(\ref{Nj1}) of the {\it open} charm hadrons will, however, be multiplied 
by the additional `canonical suppression' factor 
(see e.g. \cite{Go1}). This suppression factor
is the same for all individual single charm states. It leads to 
the total open charm multiplicity $N_O^{ce}$ in the c.e.:
\begin{equation}\label{NO}
N_O^{ce} ~=~N_O~\frac{I_1(N_O)}{I_0(N_O)}~,
\end{equation}
where $N_O$ is the total g.c.e. multiplicity of all charm and 
anticharm mesons and (anti)baryons calculated with Eq.(\ref{Nj1}) and
$I_0,I_1$ are the modified Bessel functions. 
For $N_O<<1$ one has $I_1(N_O)/I_0(N_O) \simeq N_O/2$ and, therefore,
the c.e. total open charm multiplicity is strongly suppressed in
comparison to the g.c.e. result. 
For $N_O>>1$ one finds $I_1(N_O)/I_0(N_O) \rightarrow 1$
and therefore $N_O^{ce}\rightarrow N_O$, i.e. the c.e. and the g.c.e
results coincide. 
In high energy A+A collisions the total number
of strange and antistrange hadron is much larger than unity.
Hence the strangeness conservation can be considered in
g.c.e. approach.
The same  is valid for baryonic number and electric charge. 
This is, however, not the case for the charm. 
At the SPS energies the c.e. suppression effects 
are always important:
even in the most central Pb+Pb collisions the total number of
open charm hadrons is expected to be smaller than one.
It will be shown that the c.e. treatment of charm  
conservation remains crucially important also at the RHIC energies 
for the studies of the $N_p$ dependence of the
open charm and charmonium production.
Therefore, in what follows, the baryonic
number, electric charge and strangeness of the HG system are treated
according to the g.c.e. but charm is considered in 
the c.e. formulation where the exact "charm charge" conservation 
is imposed.


Hence we formulate our model as follows.
The charm quark-antiquark pairs are assumed to be created at the early
stage of A+A collisions. In the subsequent evolution of the system,
the number of this pairs remains approximately constant and is
not necessary equal to its equilibrium value. The deviation from
the chemical equilibrium should be taken into account by  the charm 
enhancement factor $\gamma_c$.
The {\it distribution} of created $c\overline{c}$ pairs among open 
and hidden charm hadrons is regulated by statistical model
according to Eq.(\ref{Nj1}) with account for the canonical suppression
(\ref{NO}). So the statistical coalescence model in the c.e. is
formulated as:
\begin{equation}\label{SCM}
N_{c\overline{c}}^{dir}
~=~\frac{1}{2}~\gamma_c~N_O~
\frac{I_1(\gamma_c N_O)}{I_0(\gamma_c N_O)}~
+~\gamma_c^2~N_H~,
\end{equation}
where $N_H$ is the total multiplicities of hadrons with hidden charm.
(Note that the second term in 
the right-hand side of Eq.(\ref{SCM}) gives only a tiny correction 
to the first term, i.e. most of the directly created 
$c\overline{c}$ pairs are transformed
into the open charm hadrons.)
To find $N_O$ and $N_H$ we use Eq.(\ref{Nj1}) for
the individual hadron thermal multiplicities 
in the g.c.e. and take the summation over all known particles and
resonances \cite{pdg} with open and hidden charm, 
respectively\footnote{Note that possible (very small) contributions 
of particles with double open charm are neglected in Eq.(\ref{SCM}).}.

Provided that $N_O$, $N_H$ and $N_{c\overline{c}}^{dir}$ are known,
$\gamma_c$ can be found from Eq.(\ref{SCM}).
The $J/\psi$ multiplicity is then given by
\begin{equation}\label{Jpsi}
\langle J/\psi \rangle~= ~\gamma_c^2~N_{J/\psi}^{tot}~.
\end{equation}
where $N_{J/\psi}^{tot}$ is given by 
\begin{equation}\label{psitot}
N_{J/\psi}^{tot}=N_{J/\psi}~+~{\rm Br}(\psi^{\prime})N_{\psi^{\prime}}~+~
~{\rm Br}(\chi_1)N_{\chi_1}~+~
{\rm Br}(\chi_2)N_{\chi_2}~,
\end{equation}
$N_{J/\psi}$, $N_{\psi^{\prime}}$, $N_{\chi_1}$, $N_{\chi_2}$
are calculated according to Eq.(\ref{Nj1}) and
${\rm Br}(\psi^{\prime})\simeq
0.54$, ${\rm Br}(\chi_1)\simeq 0.27$, 
${\rm Br}(\chi_2)\simeq 0.14$ are the
decay branching
ratios of the excited charmonium states into $J/\psi$.

Hence, to calculate the $J/\psi$ multiplicity in SCM we need 
the following information:\\
 1) the chemical freeze-out 
(or hadronization) parameters $V,T,\mu_B$ for A+A collisions 
at high energies;\\
 2) the number
$N_{c\overline{c}}^{dir}$ of $c\overline{c}$-pairs created 
in hard parton collisions at the early stage of A+A reaction.

For the RHIC energies the chemical freeze-out temperature $T$ 
is expected to be close to that for the SPS energies: 
$T=175\pm 10$~MeV.
To fix the unknown
volume parameter $V$ and baryonic chemical
potential $\mu_B$ we use the parametrization of the total
pion multiplicity \cite{Ga:pi}:
\begin{equation}\label{pionexp}
\frac{\langle \pi \rangle}{N_p} ~\simeq~
C ~ \frac{(\sqrt{s}-2m_N)^{3/4}}{(\sqrt{s})^{1/4}}
\end{equation}
where $C = 1.46$ GeV$^{-1/2}$
and $m_N$
is nucleon mass. For the RHIC energies the nucleon mass in 
Eq.(\ref{pionexp}) 
can be neglected so that
\begin{equation}\label{pionexps}
\langle \pi \rangle  
 ~\simeq ~C~N_p~(\sqrt{s})^{1/2}.
\end{equation}
Eq.(\ref{pionexp}) is an agreement with both the SPS data and
the preliminary RHIC data in Au+Au collisions at
$\sqrt{s}=56$~GeV and $ \sqrt{s}=130$~GeV.
The pion multiplicity (\ref{pionexp}) should be equated to
the total HG pion multiplicity $N_{\pi}^{tot}$ which includes the pions
coming from the resonance decays (similar to Eq.(\ref{psitot})).
The HG parameters $V$ and $\mu_B$ are found then as the solution
of the following coupled equations:
\begin{eqnarray}
\label{pi}
\langle \pi \rangle ~ & = & N_{\pi}^{tot}(V,T,\mu_B)~\equiv~
V~n_{\pi}^{tot}(T,\mu_B)~,\\
 \label{Np}
N_p~ & = & ~V~n_B(T,\mu_B)~,
\end{eqnarray}
where $n_B$ is the HG baryonic density. In
these calculations we fix the temperature parameter $T$.
The baryonic chemical potential for Au+Au collisions
at the RHIC energies is small
 ($\mu_B < T$) and decreases with collision energy. Therefore,
most of the thermal HG multiplicities
become close to their limiting  values at
 $\mu_B \rightarrow 0$.
Consequently most of hadron ratios $N_j^{tot}/N_i^{tot}$
become independent of the collision energy. The volume of 
the system is approximately proportional to the number of 
pions: 
\begin{equation}
V \sim \langle \pi \rangle \sim N_p  (\sqrt{s})^{1/2}~.
\end{equation}

Note that $T = 170 \div 180$~MeV leads to the HG value of the thermal
ratio:
\begin{equation}\label{psiratio}
 \frac{\langle \psi^{\prime}\rangle}
{\langle J/\psi \rangle} ~=~
\left(\frac{m_{\psi^{\prime}}}{m_{J/\psi}}\right)^{3/2}~
\exp \left(-~\frac{m_{\psi^{\prime}} - m_{J/\psi}}{T}\right)~=~0.04\div
0.05~,
\end{equation}
in agreement with data \cite{ratio} in central ($N_p>100$) Pb+Pb
collisions at the CERN SPS. This fact was first noticed in Ref.\cite{Sh}.
At small $N_p$ as well as in p+p and p+A collisions the
measured value of the  $\langle \psi^{\prime} \rangle /\langle J/\psi
\rangle$ ratio is several times larger than its statistical mechanics  
estimate (\ref{psiratio}). 
Our analysis of the SCM 
will be therefore restricted
to A+B collisions with $N_p> 100$. We do not intend 
to describe the open and hidden charm production in p+p, p+A
and very peripheral A+B collisions within the SCM.

The number of directly produced
$c\overline{c}$ pairs, $N^{dir}_{c\overline{c}}$, in the left-hand
side of Eq.(\ref{SCM}) will be estimated in the Glauber approach.
For $A+B$ collision at the impact parameter $b$, this number
is given by the formula:
\begin{equation}\label{Ndir}
N^{dir}_{c\overline{c}}(b)~=~
AB~T_{AB}(b)~
\sigma(NN \rightarrow c\overline{c} + X)~,
\end{equation}  
where  
$\sigma(NN \rightarrow c\overline{c} + X)$ is the 
cross section  of $c\overline{c}$ pair production in 
$N+N$ collisions and 
$T_{AB}(b)$ is the nuclear 
overlap function (see Appendix for details).

The cross section of $c\overline{c}$ pair production in 
N+N collisions can be calculated in the pQCD.
Such calculations (in the leading order of the pQCD) were
first done in Ref.\cite{comb}. 
We use the  next-to-leading order result presented in Ref.\cite{ruusk}.
This result was obtained with GRV HO \cite{GRV} structure functions, the 
$c$-quark mass and renormalization scale were fixed at $m_c=\mu=1.3$~GeV
to fit the available experimental data.   
We parametrize the $\sqrt{s}$-dependence of the cross section
for $\sqrt{s}=20\div200$~GeV  as:
\begin{equation}\label{pert1}
\sigma(pp\rightarrow c\overline{c})~=~\sigma_0~
\left(1- \frac{M_{0}}{\sqrt{s}}\right)^{\alpha}~
\left(\frac{\sqrt{s}}{M_{0}}\right)^{\beta}~,
\end{equation}
with $\sigma_0 \approx 3.392 $~$\mu$b, $M_{0} \approx 2.984$~GeV,
$\alpha \approx 8.185$ and $\beta \approx 1.132$.
Note that free parameters of the pQCD calculations
in Ref.\cite{ruusk} were fitted
to the existing data, therefore, our parametrization (\ref{pert1})
is also in agreement with data on the total charm production
in p+p collisions.


The average number of participants (`wounded nucleons')
in A+B collisions at impact parameter $b$ is given by \cite{Bialas}
\begin{eqnarray}\label{Npart}
N_p(b) &=& A~\int_{-\infty}^{+\infty}  d x
\int_{-\infty}^{+\infty}  d y  
~T_A(\sqrt{x^2+y^2})\left[1 - \left(1-
\sigma_{N N}^{inel} T_B(\sqrt{x^2+(y-b)^2})\right)^B\right] \nonumber \\
& &  +~
B~\int_{-\infty}^{+\infty}  d x
\int_{-\infty}^{+\infty}  d y 
~T_B(\sqrt{x^2+(y-b)^2})\left[1 - \left(1-
\sigma_{N N}^{inel} T_A(\sqrt{x^2+y^2})\right)^A\right]~, 
\end{eqnarray}
where $T_A(\vec{s})$ ($T_B(\vec{s})$) is the nuclear thickness
function for the nucleus $A$ ($B$) and
$\sigma_{N N}^{inel}$ is the total inelastic
cross section of $N+N$ interaction. 
To parametrize the $\sqrt{s}$-dependence of 
$\sigma_{N N}^{inel}$ we made the assumption that
in the energy range $\sqrt{s}=10\div200$~GeV, it is
proportional to the total $NN$ cross section $\sigma_{N N}$ and use
the standard fit for $\sigma_{N N}$ \cite{pdg}:
\begin{equation}\label{siqin}
\sigma_{N N}^{inel}   
\approx
0.7 \sigma_{N N} \simeq
0.7~ (X_{pp} s^{\epsilon} +
Y_{1pp} s^{-\eta_1} - Y_{2pp} s^{-\eta_2} ),
\end{equation}
where $\epsilon=0.093$, $\eta_1=0.358$, $\eta_2=0.560$,
$X_{pp}=18.751$, $Y_{1pp}=63.58$ and $Y_{2pp}=35.46$.

\begin{figure}[p]
\begin{center}
\vfill
\epsfig{file=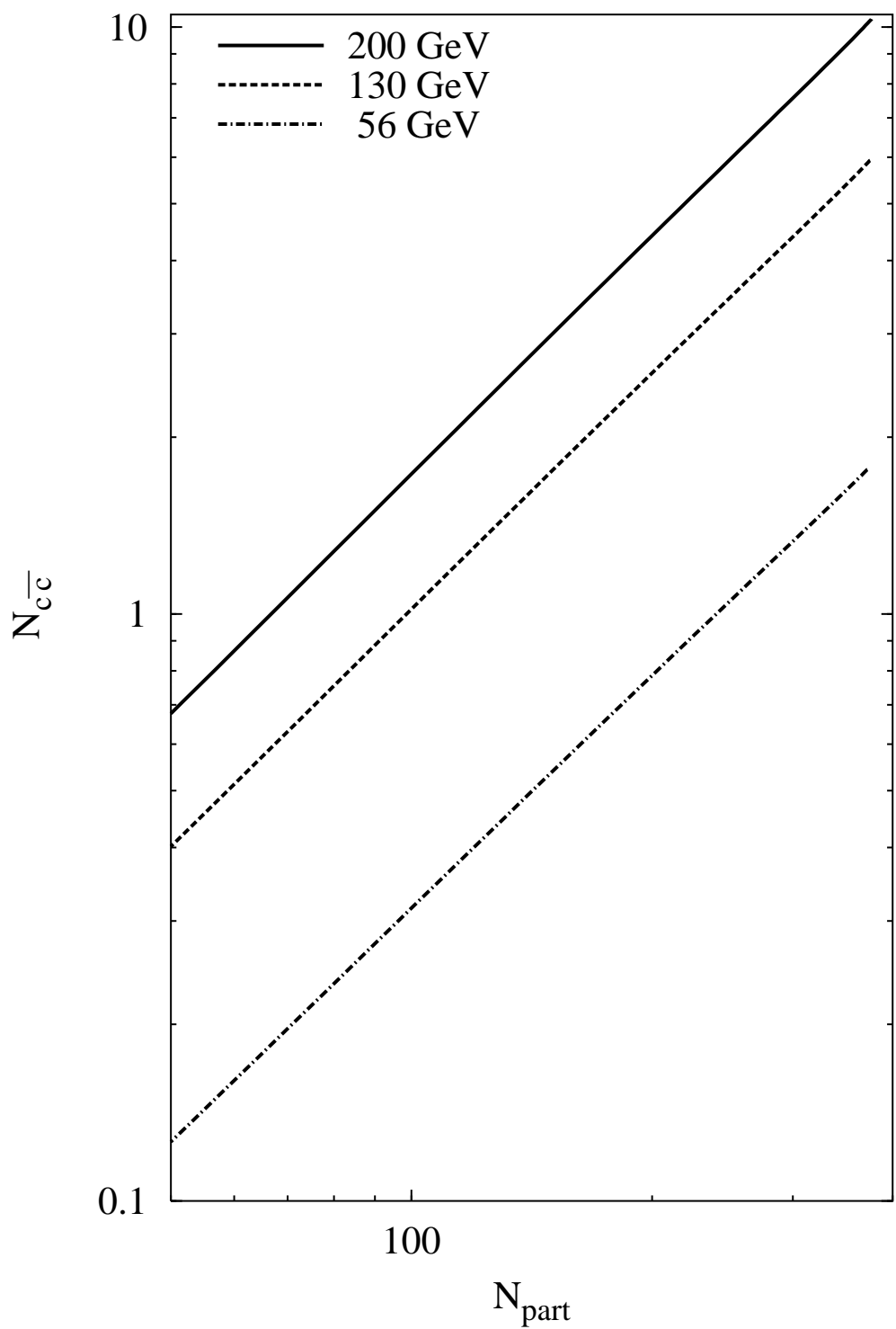,height=20cm}
\mbox{}\\
\vfill
\caption{$N_{c\overline{c}}^{dir}$ versus $N_p$
for $\sqrt{s} = 56, 130, 200$~GeV.
\label{Ncc}
}
\end{center}
\end{figure}

Eqs. (\ref{Ndir}) and (\ref{Npart}) provide
parametric dependence of the number of produced $c\overline{c}$ pairs
on the number of participating nucleons, 
$N^{dir}_{c\overline{c}}= N^{dir}_{c\overline{c}}(N_p)$, which is
shown in 
Fig \ref{Ncc} for Au+Au collisions. It is seen that  the 
dependence is represented by a straight line in the double-logarithmic
scale, so that 
\begin{equation}\label{k}
N^{dir}_{c\overline{c}}~ \sim~ (N_p)^k
\end{equation}
for $N_p > 50$.
We find that $k= 1.31 \div 1.35$\footnote{It is interesting to note that 
for the most central $A+A$ collisions ($N_p \approx 2A$), 
$N^{dir}_{c\overline{c}}$
has approximately the same dependence on the atomic weight of 
the colliding nuclei: 
$N^{dir}_{c\overline{c}}~ \sim~ A^{4/3}~\sim~(N_p)^{4/3}$.}.
Using Eq.(\ref{pert1}) 
one finds then the following behavior of $N^{dir}_{c\overline{c}}$
at high energies:
\begin{equation}
N^{dir}_{c\overline{c}}~ \sim~ (N_p)^k (\sqrt{s})^\beta~.
\end{equation}

\begin{figure}[p]
\begin{center}
\vfill
\epsfig{file=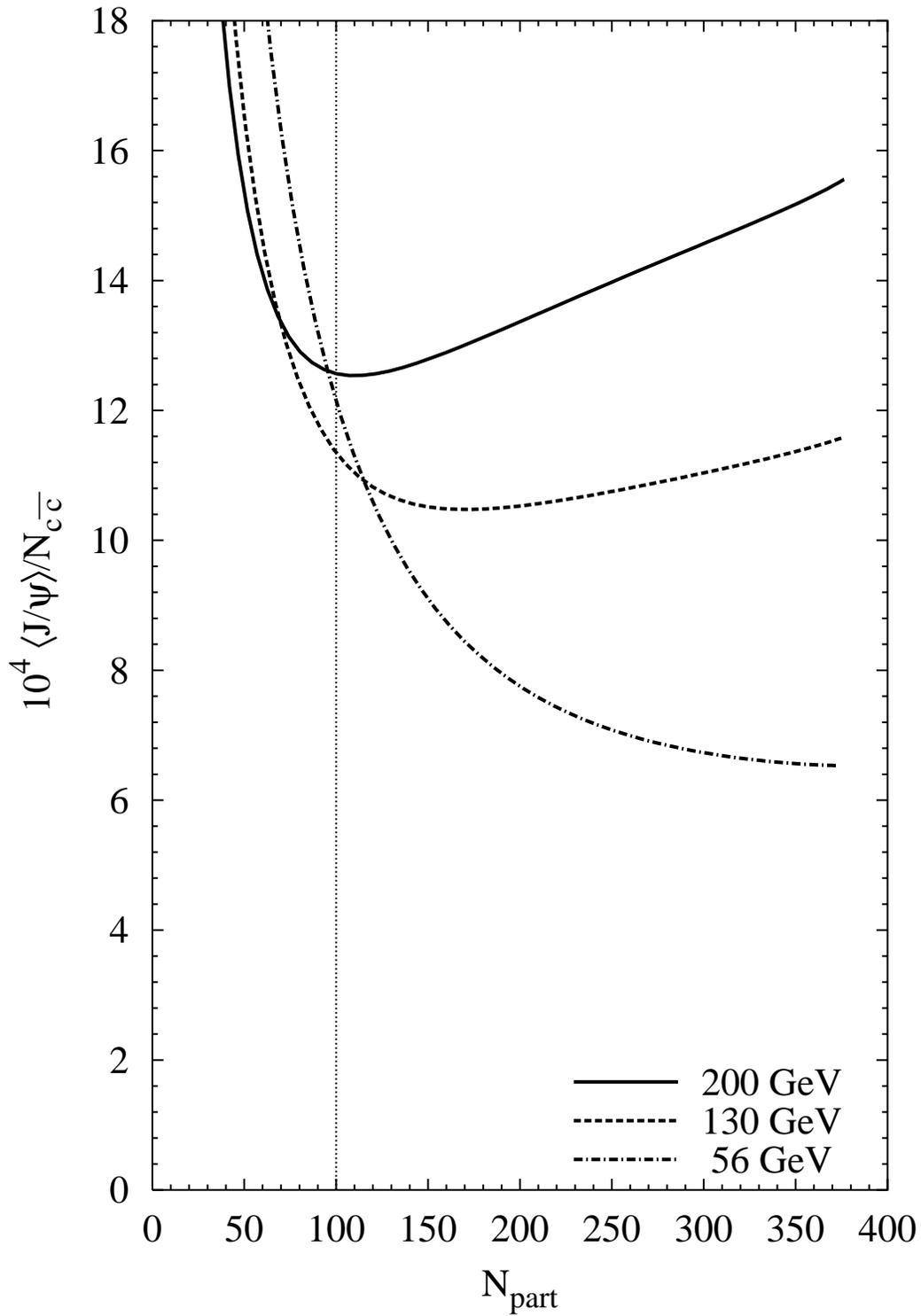,height=20cm}
\mbox{}\\
\vfill
\caption{$\langle J/\psi\rangle /N_{c\overline{c}}^{dir}$ versus
$N_p$                                                      
for $\sqrt{s} = 56, 130, 200$~GeV. The vertical line shows the 
lower bound of the applicability domain of the SCM.
\label{R_Np}
}
\end{center}
\end{figure}

Now we are able to calculate the ratio $R$ (\ref{ratio1}) in the SCM
and study its dependence on $N_p$ and $\sqrt{s}$.
The dependence of $R$ 
on the number of participants is shown in
Fig.\ref{R_Np}. It is seen that the ratio has {\it qualitatively} different 
behavior
at different energies. At the lowest RHIC energy 
$\sqrt{s}=56$~GeV, the SCM predicts decreasing of the ratio
with
the number of participants ($J/\psi$ suppression). In contrast, at the highest 
RHIC energy $\sqrt{s}=200$~GeV the ratio increases with the number of 
participant ($J/\psi$ enhancement) for $N_p>100$. 
Both suppression (at $N_p < 150$) and enhancement (at $N_p > 200$) 
are seen at the intermediate RHIC energy $\sqrt{s}=130$~GeV.

\begin{figure}[p]
\begin{center}
\vfill
\epsfig{file=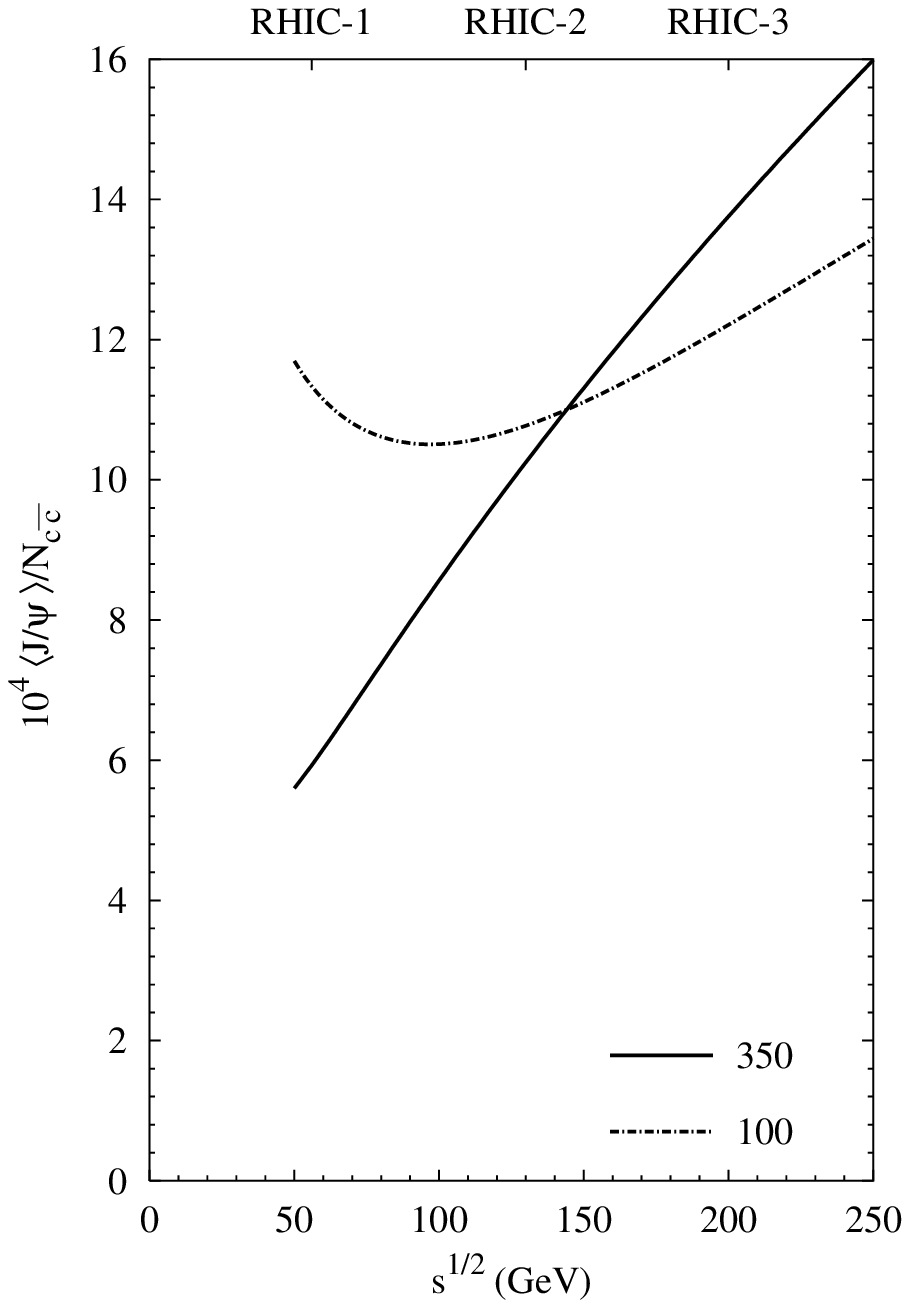,height=20cm}
\mbox{}\\
\vfill
\caption{$\langle J/\psi\rangle /N_{c\overline{c}}^{dir}$ versus
$\sqrt{s}$
for $N_p=100$ and 350.
\label{R_sqs}
}
\end{center}
\end{figure}

Similarly, there are qualitatively different dependencies of $R$ on the
collision energy for small ($N_p=100$) and for large ($N_p=350$)
number of the participants. This can be seen in Fig.\ref{R_sqs}.  
Non-monotonic dependence of the ratio $R$ on $\sqrt{s}$
is expected at $N_p=100$ .
At $N_p=350$, the ratio $R$ increases monotonically
with $\sqrt{s}$ at all RHIC energies $\sqrt{s}= 56\div200$~GeV.
The minimum of $R$ in this case corresponds to the energy region
between the SPS and RHIC: $\sqrt{s}\approx 30$~GeV.

To understand the behavior of $R$ it is instructive to study
the limiting cases: $N_{c\overline{c}}^{dir} << 1$ and
$N_{c\overline{c}}^{dir}>>1$. 
Neglecting the hidden-charm term
in Eq.(\ref{SCM}) one finds for $N_{c\overline{c}}^{dir} << 1$ :
\begin{equation}\label{Ncsm}
N_{c\overline{c}}^{dir}
~\simeq~\frac{1}{4}~\gamma_c^2~N_O^2~,
\end{equation}
hence,
\begin{equation}\label{JD1} 
R~\equiv~ \frac{\langle J/\psi\rangle} 
{N_{c\overline{c}}^{dir}}~\simeq~ \frac{
N_{J/\psi}^{tot}}
{N_O^2/4}~\sim~\frac{1}{V}~\sim ~
N_p^{-1}~ \left(\sqrt{s}\right)^{-1/2}~. 
\end{equation}
Eq.(\ref{JD1}) shows $1/V$ {\it universal} suppression of the
ratio $R$. This ratio
decreases as $N_p^{-1}$ and $\left(\sqrt{s}\right)^{-0.5}$ with 
the increasing number of participants and collision energy, and 
the shape of this $J/\psi$ {\it suppression} is essentially 
independent of the functional dependence of 
$N_{c\overline{c}}^{dir}$ on $N_p$ and $\sqrt{s}$.

If $N_{c\overline{c}}^{dir} >> 1$ one finds from Eq.(\ref{SCM}):
\begin{equation}\label{Ncla}
N_{c\overline{c}}^{dir}
~\simeq~\frac{1}{2}~\gamma_c~N_O~,
\end{equation}
so that $\gamma_c \simeq 2 N_{c\overline{c}}^{dir}/N_O
\sim N_{c\overline{c}}^{dir}/V$ and, hence,
\begin{equation}\label{JD2}
R~\equiv ~\frac{\langle J/\psi\rangle}
{N_{c\overline{c}}^{dir}}~\simeq~ \frac{\gamma_c N_{J/\psi}^{tot}}
{\gamma_c N_O/2}~\sim~\frac{N_{c\overline{c}}^{dir}}{V}~\sim ~
N_p^{k -1}~\left(\sqrt{s}\right)^{\beta - 1/2}~.
\end{equation}
According to Eq.(\ref{JD2}) the ratio $R$ increases with both $N_p$
and $\sqrt{s}$. The $J/\psi$ {\it enhancement} takes place due to the
fact that the number of primary nucleon-nucleon collisions grows faster 
than the number of participants 
($N_{c\overline{c}}^{dir} \sim (N_p)^k$, $k>1$) and because the 
pion multiplicity (and therefore the volume of the system) is less
sensitive to the  collision energy 
($\langle \pi \rangle \sim (\sqrt{s})^{1/2}$) than the 
number of $c\overline{c}$ pairs 
($N_{c\overline{c}}^{dir} \sim (\sqrt{s})^{\beta}$, $\beta > 1/2$).

It is seen from Fig.\ref{Ncc} that $N_{c\overline{c}}^{dir} << 1$ 
at the lowest RHIC energy for small numbers of participants, hence the
SCM predicts the $J/\psi$ {\it suppression}. In contrast, for the highest
RHIC energy and large $N_p$ the opposite limit  
$N_{c\overline{c}}^{dir} >> 1$ is reached. This leads to 
the $J/\psi$ {\it enhancement}. 

\vspace{0.5cm}
In conclusion,
the production of the $J/\psi$ mesons is studied in Au+Au collisions
at the RHIC energies in the statistical coalescence model
with the exact charm conservation. The
$c\overline{c}$ 
quark pairs are assumed to be created in the primary hard parton
collisions and their number is 
estimated within the pQCD. At the hadronization stage
the $c\overline{c}$ quarks  are distributed
among the open charm and
charmonium particles according to the 
hadron gas statistical mechanics in the canonical ensemble
formulation.

Decreasing of the $\langle J/\psi \rangle$ to 
$N^{dir}_{c\overline{c}}$ ratio
with increasing the number of nucleon participants $N_p$
is found at the lowest $\sqrt{s}=56$~GeV RHIC energy. At fixed 
{\it small} number of participants ($N_p \approx 100$) the ratio 
decreases with $\sqrt{s}$ between the lowest ($\sqrt{s}=56$~GeV) and the
intermediate ($\sqrt{s}=130$~GeV)
RHIC energies.
This is in a qualitative
agreement with the standard picture \cite{MS,Satz} of the $J/\psi$
{\it suppression}.
In contrast, a rise of the $\langle J/\psi \rangle$ to 
$N^{dir}_{c\overline{c}}$ ratio with the collision energy is predicted 
for central $Au+Au$ collisions. Moreover,
at the highest RHIC energy,  
the ratio is expected to grow with the number of participants, 
which is in a drastic contradiction with the standard picture.  
 The reason for this that in the standard picture the
hidden charm mesons are supposed to be created {\it exclusively} in the 
primary (hard) nucleon-nucleon collisions. It is assumed that all other 
interaction can only destroy them. 
Especially strong 
suppression of the charmonia is expected in the quark-gluon plasma
('anomalous $J/\psi$ suppression'). 
In distinction to this standard approach,
the statistical coalescence model considers
a possibility for the charmonium 
states to be formed from $c$ and $\overline{c}$ at the stage of 
the quark-gluon plasma hadronization. This possibility definitely 
cannot be ignored, when the number of produced 
$c\overline{c}$ pairs per $A+A$ collision becomes large: 
$N_{c\overline{c}}>>1$ (this happens for the central $Au+Au$ collisions
at the highest RHIC energy).
In this case the
$c$ and $\overline{c}$ pairs 
initially produced in different hard collision processes 
can recombine into
a hidden charm meson. Therefore, an increase  of the 
$\langle J/\psi \rangle$ to $N^{dir}_{c\overline{c}}$ ratio should 
be expected.
  
The hot quark gluon plasma is most probably formed at
high RHIC energies and this destroy all primarily produced
charmonium states \cite{gr:01}. However, the
hadronization of the quark gluon plasma within the SCM
reveals itself in the $J/\psi$ {\it enhancement} rather than suppression.
Another interesting phenomena may also take place: 
when the number of $c\overline{c}$ pairs becomes large,
two $c$ quarks (or two $\overline{c}$) can combine with a light 
(anti)quark and form 
a double charmed (anti)baryon. These baryons are predicted by the 
quark model but have not been observed yet. We expect that double 
(and probably triple) charmed baryons may be discovered in
the $Au+Au$ collisions at RHIC  \cite{wip}.

\vspace{0.5cm}
{\bf  Acknowledgments.}  The authors
are thankful to F. Becattini, P. Braun-Munzinger, K.A.~Bugaev,
L.~Frankfurt, M. Ga\'zdzicki, L.~Gerland, L.~McLerran,  
I.N.~Mishustin, G.C.~Nayak,  K.~Redlich and J.~Stachel for comments and
discussions.
We acknowledge the financial support of DAAD, Germany.
The research described in this publication was made possible in part 
by Award \# UP1-2119 of the U.S. Civilian Research and Development
Foundation for the Independent States of the Former Soviet Union
(CRDF).

\appendix

\section*{Nuclear geometry}

The spherically symmetrical distribution of the nucleons in 
the $Au-197$ nucleus can be parametrized by the two-parameter 
Fermi function \cite{tables} (this parametrization is also known
as Woods-Saxon distribution):
\begin{equation}\label{2pF}
\rho(r)=\rho_0 \left[1+exp\left( 
\frac{r-c}{a}
\right)\right]^{-1}
\end{equation}
with $c \approx 6.38$ fm, $a \approx 0.535$ fm and $\rho_0$ is
given by the normalization condition:
\begin{equation}\label{norm_rho}
4 \pi \int_0^\infty dr r^2 \rho(r) = 1 .
\end{equation}

The nuclear thickness distribution $T_A(b)$ is defined by the 
formula 
\begin{equation}\label{T_A}
T_A(b) = \int_{-\infty}^{\infty} d  z \rho \left( \sqrt{b^2 + z^2} \, \right)~,
\end{equation} 
and the nuclear overlap function is defined as
\begin{equation}\label{T_AB}
T_{AB}(b)=\int_{-\infty}^{\infty} d  x  
\int_{-\infty}^{\infty} d  y
T_A \left( \sqrt{x^2+y^2} \, \right) T_B \left( \sqrt{x^2+(y-b)^2} \, \right)~.
\end{equation}
 From Eq.(\ref{norm_rho}), one can deduce that 
the above functions satisfy the following normalization conditions: 
\begin{equation}\label{norm_T_AB}
2 \pi \int_0^\infty d b \, b \, T_A(b) = 1~,~~~~
2 \pi \int_0^\infty d b \, b \, T_{AB}(b) = 1 .
\end{equation}

\begin{thebibliography}{99}

\bibitem{MS}
T.~Matsui and H.~Satz,
Phys.\ Lett.\ B {\bf 178} (1986) 416.

\bibitem{Satz}
H.~Satz,
Rept.\ Prog.\ Phys.\ {\bf 63} (2000) 1511
[hep-ph/0007069].

\bibitem{comover}
C.~Spieles, R.~Vogt, L.~Gerland, S.~A.~Bass, M.~Bleicher, 
H.~Stocker and W.~Greiner,
nonequilibrium approach,''
Phys.\ Rev.\ C {\bf 60} (1999) 054901
[hep-ph/9902337];\\
J.~Geiss, C.~Greiner, E.~L.~Bratkovskaya, W.~Cassing and U.~Mosel,
Phys.\ Lett.\ B {\bf 447} (1999) 31
[nucl-th/9803008];\\
N.~Armesto, A.~Capella and E.~G.~Ferreiro,
Phys.\ Rev.\ C {\bf 59} (1999) 395
[hep-ph/9807258];\\
D.~E.~Kahana and S.~H.~Kahana,
Prog.\ Part.\ Nucl.\ Phys.\ {\bf 42} (1999) 269.

\bibitem{Gavai}
R.~Gavai, D.~Kharzeev, H.~Satz, G.~A.~Schuler, K.~Sridhar and R.~Vogt,
Int.\ J.\ Mod.\ Phys.\ A {\bf 10} (1995) 3043
[hep-ph/9502270].

\bibitem{GG}
M.~Gazdzicki and M.~I.~Gorenstein,
Phys.\ Rev.\ Lett.\ {\bf 83} (1999) 4009
[hep-ph/9905515].

\bibitem{Ka:00}
S.~Kabana,
hep-ph/0010246.

\bibitem{Br1}
P.~Braun-Munzinger and J.~Stachel,
Phys.\ Lett.\ B {\bf 490} (2000) 196
[nucl-th/0007059].

\bibitem{Go:00}
M.I. Gorenstein, A.P.~Kostyuk,
H.~St\"ocker and  W.~Greiner,
hep-ph/0010148, Phys. Lett. B (in print) and hep-ph/0012015,
J. Phys. G (in print).

\bibitem{Le:00}
P. Csizmadia and P. L\'evai,
hep-ph/0008195 (2000);\\
P. L\'evai, T.S. Bir\'o, P. Csizmadia, T.~Cs\"org\H o and
J.~Zim\'anyi,
hep-ph/0011023.

\bibitem{Ra:00}
R.~L.~Thews, M.~Schroedter and J.~Rafelski,
hep-ph/0007323.

\bibitem{Mc:00}
M.I. Gorenstein, A.P.~Kostyuk, L. McLerran,
H.~St\"ocker and  W.~Greiner,
hep-ph/0012292.

\bibitem{Ri}
D.H. Rischke, M.I. Gorenstein, H.St\"ocker
and  W.Greiner,
 Z. Phys. C {\bf 51} (1991) 485.

\bibitem{YG}
M.I. Gorenstein, G.Yen, W.Greiner and S.N.Yang,
Phys. Rev. C {\bf 56} (1997) 2210.\\
M.I. Gorenstein, A.P. Kostyuk and Ja.~Krivenko, J. Phys. G {\bf25}
(1999) L75.


\bibitem{HG} 
P.~Braun-Munzinger, I.~Heppe and J.~Stachel,
Phys.\ Lett.\ B {\bf 465} (1999) 15
[nucl-th/9903010];\\
F.~Becattini, J.~Cleymans, A.~Keranen, E.~Suhonen and K.~Redlich,
hep-ph/0002267;\\
G.~D.~Yen and M.~I.~Gorenstein,
Phys.\ Rev.\ C {\bf 59} (1999) 2788
[nucl-th/9808012].

\bibitem{Ra:91}
J.~Rafelski,
Phys.\ Lett.\ B {\bf 262} (1991) 333.

\bibitem{ratio}
M.~Gonin {\it et al.},
PRINT-97-208   
{\it Presented at 3rd International Conference on Physics and 
Astrophysics of Quark Gluon Plasma (ICPAQGP 97), Jaipur, India, 
17-21 Mar 1997}.

\bibitem{Sh}
H.~Sorge, E.~Shuryak and I.~Zahed,
Phys.\ Rev.\ Lett.\ {\bf 79} (1997) 2775
[hep-ph/9705329].

\bibitem{Go1}
K.~Redlich and L.~Turko,
Z.\ Phys.\ C {\bf 5} (1980) 201;\\ 
J.~Rafelski and M.~Danos,
Phys.\ Lett.\ B {\bf 97} (1980) 279;\\
J.~Cleymans, K.~Redlich and E.~Suhonen,
Z.\ Phys.\ C {\bf 51} (1991) 137;\\
M.~I.~Gorenstein, M.~Ga\.zdzicki and W.~Greiner,
Phys.\ Lett.\ B {\bf 483} (2000) 60
[hep-ph/0001112].

\bibitem{pdg} 
Particle Data Group,
Eur.\ Phys.\ J.\  {\bf C15} (2000) 1.

\bibitem{Ga:pi}
M. Ga\.zdzicki, private communication.

\bibitem{GaJ}
M.~Ga\.zdzicki,
Phys.\ Rev.\ C {\bf 60} (1999) 054903
[hep-ph/9809412].

\bibitem{comb}
B.~L.~Combridge,
Nucl.\ Phys.\ B {\bf 151} (1979) 429.

\bibitem{ruusk}
P.~L.~McGaughey, E.~Quack, P.~V.~Ruuskanen, R.~Vogt and X.~Wang,
Int.\ J.\ Mod.\ Phys.\ A {\bf 10} (1995) 2999
[hep-ph/9411438].

\bibitem{GRV}
M.~Gluck, E.~Reya and A.~Vogt,
Z.\ Phys.\ C {\bf 53} (1992) 127.

\bibitem{Bialas}
A.~Bialas, M.~Bleszynski and W.~Czyz,
Nucl.\ Phys.\ B {\bf 111} (1976) 461.



\bibitem{gr:01} L. Grandchamp and R. Rapp,
hep-ph/01033124.

\bibitem{wip}
A.P.~Kostyuk et al, work in progress.

\bibitem{tables}
C.~W.~De Jager, H.~De Vries and C.~De Vries,
Atom.\ Data Nucl.\ Data Tabl.\ {\bf 14} (1974) 479.

\end{thebibliography}
\end{document}